\documentclass[12pt]{iopart} 
\usepackage{epsfig,multicol,iopams,graphicx}

\def \pre{Preprint}
 
\def \w {\omega} 
\def \I {\mathfrak {I}}

\def \M {\mathcal {M}} 

\begin{document}
\title{Entropy spectra of black holes from resonance modes in scattering by the black holes}

\author{Yongjoon Kwon and Soonkeon Nam}

\address{ Department of Physics and Research Institute of Basic Sciences, \\
 Kyung Hee University, Seoul 130-701, Korea}

\eads{\mailto{emwave@khu.ac.kr}, \mailto{nam@khu.ac.kr}}



\begin{abstract}
Since the Bekenstein's proposal that a black hole has equally spaced area spectrum, the quasinormal modes as the characteristic modes of a black hole have been used  in obtaining the horizon area spectrum of the black hole. However, the area spectrum of the Kerr black hole in some previous works was inconsistent with the Bekenstein's proposal. 
In this paper, noting that black holes can have three types of resonance modes which are quasinormal modes (QNM), total transmission modes (TTM), and total reflection modes (TRM), we propose that all of these modes in highly damped regime should be used in quantizing the black hole. Although the QNM and the TTM of the Kerr black hole give us complicated quantization conditions from the Bohr-Sommerfeld quantization of action variable, we find a very simple result from the TRM. It gives equally spaced outer horizon area.  
Therefore by Bekenstein-Hawking area law we find that the Kerr black hole has universal behavior of the equally spaced entropy spectrum. 
With the same argument, we find that the Reissner-Nordstr{$\rm {\ddot {o}}$}m black hole also has equally spaced entropy spectrum.
\end{abstract}
%


\pacs{04.70.Dy, 
04.70.-s, 
}

\maketitle

\section{Introduction}
%
As a quantum property of black holes, it is believed that the black hole horizon area is quantized. 
It was first proposed by Bekenstein \cite{be} that the horizon area is an adiabatic invariant and should be quantized, based on Ehrenfest principle  that any classical adiabatic invariant corresponds to a quantum entity with discrete spectrum. By considering the minimum change of the horizon area in the process of the assimilation of a test particle into a black hole, it was obtained that the area spectrum should be linearly quantized, i.e. $A= \gamma n \hbar$ where $\gamma$ is an undetermined dimensionless constant  \cite{be}. 
However there are other approaches which suggest more complicated area spectrum \cite{1302} and some attempts to make a point of contact with Bekenstein's idea via proper interpretation \cite{122dryer}. 

When a perturbation on a black hole is given, the black hole undergoes damped oscillations which are called  quasinormal modes. 
By using these quasinormal modes of a black hole, it was realized that the area spectrum of the black hole can be obtained in the semiclassical limit \cite{hod, kun, setvag, magi, vage, med, wei, wei2}. 
By considering the real part of the asymptotic quasinormal modes of a black hole as a transition frequency in the semiclassical limit, the area spectrum of the Schwarzschild black hole was obtained as $A = (4\, {{\rm ln} 3}) \, n \hbar$  \cite{hod}.  
However  Hod's original conjecture \cite{hod} was shown to be inconsistent  for other black holes \cite{020, 173, 029}. 
The area spectrum of the Schwarzschild black hole was reproduced  by considering an adiabatic invariant of the system with energy $E$ and vibrational frequency $\w(E)$, given by  the real part of the asymptotic quasinormal modes,  via the Bohr-Sommerfeld quantization \cite{kun}. 
For a rotating black hole, a modified form of adiabatic invariant was suggested as  $\mathcal I =\int {{dE-\Omega dJ} \over \w(E)} = n \hbar$ via  the Bohr-Sommerfeld quantization \cite{setvag}. 
Later, it was proposed that a perturbed black hole should be described as a damped harmonic oscillator and a transition in a black hole should be considered as the transition between quantum levels $(\w_{0})_k\equiv ( \sqrt{\w_{R}^2 +\w_I^2 } )_k$, where $\w_R$ and $\w_I$ are the real and imaginary parts of the asymptotic quasinormal modes \cite{magi}. Therefore the characteristic classical frequency $\w_c$  should be identified with the transition frequency between the quantum levels $(\w_{0})_k$  in the semiclassical limit, i.e. $\w_{c} = (\w_ {0})_{k} - (\w_ {0})_{k-1} \simeq ( | \w_ {I} | )_{k} - ( | \w_ {I} | )_{k-1} $ for the highly damped quasinormal modes, where $k \in N , k \gg 1$  \cite{magi}. 
By using this transition frequency $\w_c$, the area spectrum of the Schwarzschild black hole was obtained  as $A=8 \pi n \hbar $ \cite{magi, med, wei}. 
For the Kerr black hole, the asymptotic quasinormal modes have been first investigated numerically in Refs.\cite{029, 052}.
The transition frequency $\w_c$  can be obtained from the asymptotic quasinormal modes which are analytically calculated as 
\begin{equation} \label{keqnm2}
\w = \tilde \w _0 - i 4 \pi T_0 (a) \left(n+{1 / 2} \right) ~,
\end{equation}
where $n$ is integers, $\tilde \w _0 $ is a function of the black hole parameters whose real part asymptotically approaches ${\rm Re}(\tilde \w_0) \propto m$, and  $T_0 (a\equiv J/M) \approx - T_H (a=0)/2 $ within $\sim3\%$ accuracy \cite{keshk, keshk2}. The explicit expression of the imaginary part of the quasinormal modes is given in terms of the elliptic integrals \cite{4195}, and it is too complicated to calculate the area spectrum by using adiabatic invariant.  
In the previous works   \cite{vage, med} on the area spectrum of the Kerr black hole, with the transition frequency   approximately taken as $\w_c \approx  2 \pi T_H (a=0) =1/(4 M)$, the quantization of the modified adiabatic invariant, i.e. $\mathcal I =\int {{dE-\Omega dJ} \over \w_c} = n \hbar$, was calculated. But this gives  non-equally spaced area spectrum for the Kerr black hole, which is inconsistent with Bekenstein's proposal. 
Only for slowly rotating case with small angular momentum $J$ compared to  mass  $M$  of the Kerr black hole, it was found that the area spectrum was approximately equally spaced as $A= 8 \pi n \hbar$ \cite{med, myung}. 
In our recent work \cite{sk} it was reminded that an action variable is adiabatic invariant, but not every adiabatic invariant is an action variable, and that only action variable can be quantized via the Bohr-Sommerfeld quantization in the semiclassical limit.  Therefore not an adiabatic invariant but an  action variable of the classical system  should be identified in order to  apply  the Bohr-Sommerfeld quantization  \cite{sk}. 
By Bohr's correspondence principle which says that the transition frequency at large quantum number equals to the classical oscillation frequency of the corresponding classical system, a black hole with the transition frequency $\w_c$ can be considered  as  the  classical system of periodic motion with oscillation frequency $\w_c$ in the semiclassical limit. Therefore the action variable of the classical periodic system with the oscillation frequency $\w_c$ was identified and finally quantized via the  Bohr-Sommerfeld quantization in the semiclassical limit as follows \cite{sk}:
\begin{eqnarray} \label{myf}
{\I} =\int { dE \over {\w _c }} =\int { dM \over {\w _c }} = n \hbar ~~, ~~(n \in Z, \vert n \vert \gg 1)~,
\end{eqnarray}
where   the transition frequency $\w_c$ in the semiclassical limit is given by $\w_{c} = ( | \w_ {I} | )_{k} - ( | \w_ {I} | )_{k-1} $ for highly damped modes, and the change of the energy $E$ of a black hole is considered as the change of the ADM mass $M$ (or ADT mass $\M$ according to the gravity theory \cite{sk2}). 
This formula can be also applied for a rotating  black hole with a transition frequency. 
For example, the area and entropy spectra of the BTZ and  warped ${\rm AdS}_3$ black holes were obtained in Refs.\cite{sk, sk2}, where it was found that  there is the universality that the entropy spectrum of a black hole is equally spaced, even though area spectrum is not equally spaced \cite{sk2}. 

In this paper, we would like to apply the formula (\ref{myf}) for the Kerr black hole and to find if the entropy  has the universal behavior of equally spaced spectrum.  
Until now, in spite of the several attempts \cite{setvag, vage, med}  for the area spectrum of the Kerr black hole, the results  were not consistent with Bekenstein's original proposal. 
Recently, by considering the scattering problem on the Kerr black hole the highly damped quasinormal modes (QNM) of the Kerr black hole, which are given by Eq.(\ref{keqnm2}), were obtained from the poles of the transmission and reflection amplitudes \cite{keshk2}. 
However we notice that there are other resonance modes of the Kerr black hole, which are total transmission modes (TTM) and total reflection modes (TRM). These special modes of black holes were first considered by Chandrasekhar \cite{chan}, and more investigated in some works \cite{ono, van, 4306, 29751025}. 
The TRM and TTM  can be obtained from the zeros of the transmission and reflection amplitudes, respectively. 
In order to find out the property of a black hole, we have to do scattering experiment on the black hole. 
In scattering problem of a black hole, the wave equation becomes Schr${\rm \ddot o}$dinger-like equation in quantum mechanics. Then the incident waves from infinity are reflected and transmitted because of the effective potential which plays a role of potential barrier in quantum mechanics. 
So, when we consider  scattering problem on a black hole, the black hole can have other resonance modes of TTM and TRM as well as QNM. 
The QNM only depend on the black hole parameters like mass, charge, and angular momentum  which characterize a black hole.   
Therefore it has been considered that QNM are the characteristic modes of the black hole as a fingerprint in directly identifying the existence of a black hole \cite{nolkz} and  carry some information about quantum structure of  the black hole \cite{magi}. More specifically, it was proposed that the imaginary part of the highly damped quasinormal modes  represents the energy levels as quantum structure of black hole in the semiclassical limit, so that the transition frequency between the quantum levels is corresponding to the energy of emitted quanta from black hole \cite{magi}. 
We notice that TTM and TRM  are also  the characteristic modes of a black hole, since they only depend on black hole parameters. In this sense, we propose that the highly damped TTM and TRM also have the quantum structure of a black hole as the highly damped QNM. 
Therefore we should promote  TTM and TRM to the equivalent position of QNM, so that they play the same role as QNM in quantizing 
a black hole. 
Then  we can obtain the transition frequencies corresponding to each of all of them in the semiclassical limit. 
By considering other resonance modes for the Kerr black hole, we will find that the area spectrum  is consistent with Bekenstein's proposal \cite{be} and the entropy spectrum has the universality  of equally spaced spectrum. Moreover, with the same argument, we will also find that the Reissner-Nordstr{$\rm {\ddot {o}}$}m black hole has the same behavior in area and entropy spectra. 
Throughout this paper, the Planck units with $c=G=\hbar=1$ are used.

\section{Resonance modes of black hole in greybody factor and Hawking radiation }

In this section, we will briefly review some features of resonance modes of black holes. 
The perturbations of black hole spacetimes are represented by the radial Schr$\ddot {\rm o}$dinger-like wave equations of the form 
\begin{equation} \label{weq}
 { {\partial ^2_z f(z)} } + \left(\w^2 -V_z(z) \right) f(z) =0  ,
\end{equation}
where $z=z(r)$ is a tortoise coordinate which has the behavior of 
$z \sim r $ as $r \rightarrow \infty$ and $z \rightarrow - \infty$ as $r \rightarrow r_+$ with  outer horizon radius $r_+$. 
The scattering problem for  the incident wave  from spatial infinity gives the wavefunction as the solution of the above wave equation, which satisfy the following boundary conditions:
\begin{eqnarray}
f(z)_{\w} \sim \cases {  e^{- i \w z} +R(\w) e^{i \w z}   ~~~, ~~~ as~z \rightarrow \infty \\
  T(\w) e^{-i \w z}   ~~~, ~~~ as~z \rightarrow  - \infty }
\end{eqnarray}
where $T(\w)$ and $R(\w)$ are, respectively, the transmission and reflection amplitudes, and the purely ingoing wave at horizon is imposed as a boundary condition. 
We can also consider the wavefunction for $-\w$ which solves the wave equation (\ref{weq}) ;
\begin{eqnarray}
f(z)_{-\w}  \sim \cases { e^{i \w z} +\tilde R(-\w) e^{-i \w z}   ~~~, ~~~ as~z \rightarrow \infty \\
 \tilde T(-\w) e^{i \w z}   ~~~, ~~~ as~z \rightarrow  - \infty }
\end{eqnarray}
where  $\tilde T(-\w)$ and $\tilde R(-\w)$ are some other transmission and reflection amplitudes. 
Then the conserved flux is given by
\begin{equation}
\mathcal F= {1 \over {2 i}} \left( f(z)_{-\w} \partial_z f(z)_{\w} - f(z)_{\w} \partial_z f(z)_{-\w}  \right) .
\end{equation}
By calculating the conserved flux at both limits of $z$, we obtain the following relation:
\begin{equation}
T(\w) \tilde T(-\w)  + R(\w) \tilde R(-\w)  =1 ~,
\end{equation}
where $T(\w) \tilde T(-\w)$ represents absorption (transmission) probability which is associated with greybody factor of a black hole. 
By considering the scattering problem on a black hole, we can find the resonance modes of a black hole, i.e. TTM, TRM, and QNM, from the transmission and reflection probabilities. The TTM and TRM are respectively obtained from the zeros of ${R(\w) \tilde R(-\w)}$ and ${T(\w) \tilde T(-\w)}$,  while the QNM corresponds to  the poles of them. 

In order to grasp some physical meaning of the resonance modes in some way, let us point out that there is the relation between Hawking radiation and the greybody factor with the information of the resonance modes of a black hole.   
It is well known that a black hole emits Hawking radiation which has the spectrum of the blackbody radiation \cite{area0}. The decay rate of a black hole at event (outer) horizon is given by \cite{area0}
\begin{equation}
\Gamma(\w)={1 \over {e^{ { (\w- m \Omega)} /T_H}  \mp 1} }  \equiv n_H(\w) .
\end{equation}
The minus(plus) sign corresponds to bosons(fermions). 
This Hawking formula for emission spectrum indicates that the black hole is a thermal object.
To a static observer at spatial infinity, however, the spectrum of Hawking radiation is not thermal \cite{neitz}.
Since the curvature of the spacetime geometry outside event horizon plays a role of potential barrier, it filters Hawking radiation in such way that some of radiation are transmitted to infinity and the rest are reflected into black hole. Therefore the decay rate at infinity is  given by multiplying  Hawking radiation at horizon by a factor, i.e. so called "greybody factor" which is dependent on frequency, as follows:
\begin{equation} \label{decay2}
\Gamma(\w)=\gamma(\w)  n_H(\w) ={\gamma(\w) \over {e^{ { (\w- m \Omega)} /T_H} \mp 1} } .
\end{equation}
It means that Hawking radiation of black hole does not `blackbody' radiate, but `greybody' radiate, when it is measured at infinity. 
When we consider the scattering problem where the incident wave originates from infinity, the greybody factor is defined as the absorption (transmission) probability of a black hole. Therefore the decay rate of Eq.(\ref{decay2}) is given by
\begin{equation} \label{decay3}
  \qquad  \Gamma(\w)={\gamma(\w) \over {e^{ { (\w- m \Omega)}/ T_H} \mp 1} } ={ {T(\w) \tilde T(-\w)} \over {e^{ { (\w- m \Omega)} /T_H} \mp 1} } . 
\end{equation}
Therefore from Eq.(\ref{decay3}) we find that the TTM is the resonance modes which give the decay rate in the form of blackbody radiation even at infinity. 
For the TRM,  it seems that Eq.(\ref{decay3}) vanishes since the TRM corresponds to ${T(\w) \tilde T(-\w)}=0$. However, it is pointed out in Ref.\cite{keshk2} that the greybody factor has zeros only where $n_H$ has poles in general, in particular for the spherical black holes in Ref.\cite{neitz}.  Therefore the black holes we will consider in this paper have non-vanishing emission spectrum $\Gamma(\w)$ which is not affected by the TRM. In other words, the TRM is corresponding to the poles of the spectrum of Hawking radiation measured at outer horizon. In this sense, the TRM is associated with the outer horizon. On the other hand, the QNM is corresponding to the  poles of spectrum of Hawking radiation measured at infinity.

\section{Area and entropy spectra from TRM}
For the quantization of the Kerr black hole, we will consider all of the resonance modes (i.e. TTM, TRM, and QNM) and apply them to the formula (\ref{myf}). 
Before that, let us consider the Schwarzschild black hole case first as a warm-up exercise. 
In the highly damped regime, the transmission and reflection probabilities are given by \cite{neitz} 
\begin{equation}
\fl T(\w) \tilde T(-\w)  \approx { {e^{{\w} \over T_H^s} -1} \over {e^{{\w} \over T_H^s} +(1+2 \cos {\pi j}) } } ~~ {\rm and}  ~~ R(\w) \tilde R(-\w)  \approx { {2 (1+\cos{\pi j}) } \over {e^{{\w} \over T_H^s}+(1+2 \cos {\pi j})} } .
\end{equation}
The scalar, electromagnetic and gravitational perturbations correspond to $j=0$, $j=1$ and $j=2$, respectively. Since the effective potential in Schr$\ddot {\rm o}$dinger-like wave equation has the term of $(1-j^2)/r^4$, it becomes singular for the electromagnetic perturbation ($j=1$)  \cite{neitz}. Therefore the above results hold except for that case. 
For the scalar and gravitational perturbations, we can find QNM and TRM of  the Schwarzschild black hole, but there is no TTM. The QNM and TRM for the  highly damped modes are easily obtained as follows:
\begin{eqnarray} 
\w^{QNM} &=&  T_H ^s  {\rm ln} 3 -i 2 \pi  T_H ^s  \left(k+1/2 \right) ,   \\
\w^{TRM} &=& -i 2 \pi  T_H ^s  k  ,    \quad  ( k \in N ~{\rm and}~  k \gg 1 )
\end{eqnarray}
where $ T_H ^s$ is the Hawking temperature of the Schwarzschild black hole, and we take the time dependence of the wavefunction   as $e^{-i \w t}$. 
We find that the TRM gives the same transition frequency of $\w_c = 2 \pi  T_H ^s  =1/(4 M)$ as one from the QNM.  
In most cases,  QNM was enough in finding the horizon area spectrum of a black hole. But for some cases it is not sufficient. In particular for black holes with two horizons,  other resonance modes  may be needed to obtain two transition frequencies of a black hole, and from which the spectra of the both inner and outer horizon areas can be obtained. In our previous works \cite{sk, sk2}, for example, we have seen that  the spectra of the both inner and outer horizon areas are obtained from the two transition frequencies which are read off from two families of the QNM. 

The linearized and massless perturbation of the Kerr black hole is described by Teukolsky's equation \cite{teu}. 
The wave equation for the Kerr black hole can be solved in the highly damped regime by using WKB approximation along specific contours in the complex $r$-plane \cite{keshk2}.  
Using the monodromy matching method along two different contours,  the transmission and reflection probabilities of the Kerr black hole can be obtained  \cite{keshk2}. Three resonance modes, which are QNM, TTM and TRM, correspond to the poles and zeros of them. 
The resonance modes in the highly damped regime has the form of 
\begin{equation}
\w^j = \tilde \w ^j + i 4 \pi T^j  \left(n+{\mu^j / 4} \right) ,
\end{equation}
where $n$ is integers, $\mu^j $ are Maslov indices and $j$ denotes  the three resonance modes. 
The  highly damped QNM for  the wavefunction with the time dependence of   $e^{-i \w t}$  is given by   \cite{keshk2}
\begin{equation}
\w^{QNM} = \tilde \w _0 + i 4 \pi T_0  \left(k+{1 / 2} \right) ,  \quad  ( k \in N ~{\rm and}~  k \gg 1 ) , 
\end{equation}
where $\tilde \w^{QNM} \equiv \tilde \w_0$ is a function of black hole parameters. 
Note that  $T_0$ is negative vlaue.
The $|T_0|$ is a  monotonically increasing function of $a \equiv J/M$ and  $T_0 (a) \approx - T_H (a=0)/2 $ within about 3\% accuracy with $T_0 (a \rightarrow 0)= - T_H(a=0)/2 $. 
In Ref.\cite{4195}, it is shown that the explicit expression of the quasinormal modes can be given in terms of the elliptic integrals. 
The TTM and TRM  can be obtained from the exact relations between the parameters $T^j$ and $\tilde \w^j$  of the three resonance modes \cite{keshk2};
\begin{eqnarray} \label{rel1}
{1 \over {2 T^{TTM}}} - {1 \over {2 T^{QNM}}} ={1 \over {2 T^{TRM}}} &=& {1 \over  T_H}  ,\\
\label{rel2}
{ {\tilde \w^{TTM}} \over {2 T^{TTM}}} - { {\tilde \w^{QNM}} \over {2 T^{QNM}}} = { {\tilde \w^{TRM}} \over {2 T^{TRM}}} &=& {{m \Omega} \over T_H} +i 2 \pi s ,
\end{eqnarray}
where $\Omega$ is the angular velocity at horizon, $T_H$ is the Hawking temperature of the Kerr black hole and $s $ is the spin 
 of the fields, i.e. gravitational ($s$= $-2$), electromagnetic ($s$= $-1$), and scalar ($s$= $0$) fields. 
Note that  $ T^{TTM}$ and $T^{TRM}$ are positive and $ T^{QNM}$ is negative \cite{keshk2}. 
The origin of this relation can be most clearly understood when we consider the specific anti-stokes lines, along which the solutions of the wave equation are purely oscillatory, associated with each of the resonance modes. 
For the quantization of the Kerr black hole, we would like to consider TRM and TTM as well as QNM in using the formula (\ref{myf}). 
However the expressions of $T^{QNM}$ and  $T^{TTM}$ are  given in very complicated forms with the elliptic integrals \cite{4195}.  These give some difficulty in calculating the action variables in Eq.(\ref{myf}). 
Nevertheless, we  find the area spectrum of the Kerr black hole from the other resonance modes, i.e. TRM. 
The interesting thing is that we simply get the highly damped TRM of the Kerr black hole from the relations (\ref{rel1}) and (\ref{rel2}) as follows:
\begin{equation}
\w^{TRM} = m \Omega - i 2 \pi T_H(k-s),   \quad  ( k \in N ~{\rm and}~  k \gg 1 ) .
\end{equation}
Therefore the transition frequency is given by
\begin{eqnarray}
\w^{TRM}_c = 2 \pi T_H = \frac{\sqrt{M^4-J^2}}{2 M  \left(M^2+\sqrt{M^4-J^2}\right) }.
\end{eqnarray}
With this transition frequency, we obtain the quantization condition from the formula (\ref{myf}); 
\begin{eqnarray} 
 \I^{TRM} = \int {{dM} \over  {\w^{TRM}_c }} = M^2+\sqrt{M^4-J^2} = n_r \hbar ,
\end{eqnarray}
where  $n_r$ are positive integers. 
Therefore, we find that the  outer horizon area are quantized as follows: \begin{eqnarray}\label{skerr4}
 A_{out}= 8 \pi n_r \hbar ~.
\end{eqnarray}
By Bekenstein-Hawking area law \cite{area0, area}, this implies that the entropy spectrum is also equally spaced as $\triangle S =2 \pi$. 
We find that unlike other black hole cases with single horizon \cite{magi, wei, wei2}, which were studied before, where the QNM gives equally spaced outer horizon spectrum, in the case of the Kerr black hole with two horizons the TRM gives the equally spaced outer horizon spectrum. From the relation (\ref{rel1}) for the gaps in the imaginary parts of the resonance modes which are associated with the transition frequencies of the Kerr black hole, we can find that the sum of action variables for TTM and QNM is equal to the action variable for TRM, which gives the quantization of the outer horizon area. 
In other words, we obtain the following relation:
\begin{eqnarray} 
 \I^{TTM} + \I^{QNM} =  \I^{TRM} =  {A_{out} \over {8 \pi}} = n_r \hbar
\end{eqnarray}
On account of the difficulty in calculations of action variables for TTM and TRM, it is not clear what quantities the TTM and QNM quantize. For the slowly rotating case, however, the roles of TTM and QNM become clearer. In that case, we can take $ T^{QNM} \simeq - T_H(a=0)/2 = -{{T_H^s} / 2} $, where $T_H^s$ is the Hawking temperature of the Schwarzschild black hole. Using  the  relations  (\ref{rel1}) and   (\ref{rel2}) with this, we find the TTM and TRM,  and from which the transition frequencies are given by
\begin{eqnarray}
\w^{QNM}_c = 2 \pi { T_H^s},~{\rm and}~\w^{\pm}_c =  \pm 4 \pi T^{\pm}
\end{eqnarray}
The plus(minus) sign denotes TRM(TTM), and  $T^{+} \equiv T_H/2$ and $T^{-} \equiv T_{in}/2 $ where $T_{in}= {\kappa_-} / (2 \pi)=  { { T_H T_H^s} / ( {T_H -T_H^s} ) }$ with the negative surface gravity $\kappa_-$ at inner horizon  \cite{muin}. 
Therefore  we find that the action variables for three resonance modes are proportional to horizon areas as follows:
\begin{eqnarray} \label{kerraction}
 \I^{QNM} = { {A_{+}+A_{-}} \over  { 8 \pi}} ,~
 {\rm and} ~
 \I^{\pm} =  \pm{ A_{\pm} \over  { 8 \pi}} .
\end{eqnarray}
 where $A_{+} \equiv A_{out}$ and $ A_{-} \equiv A_{in}  $. 
Therefore we find that the spectra of the both inner and outer horizon areas are equally spaced as $ \triangle A_{out} =\triangle A_{in} =8 \pi \hbar$. 
By Bekenstein-Hawking area law \cite{area0, area}, the entropy spectrum is also equally spaced; $\triangle S =2 \pi$. 
Therefore, for slowly rotating case we find  that  while the quantization of the action variable for QNM is associated with the quantization of the total horizon area, the TTM and TRM  lead to the quantizations of the inner horizon area and the outer horizon area, respectively. 
Therefore for black holes with multiple horizons, we conjecture that TRM rather than QNM is associated with the quantization of the outer horizon  area. 
 Indeed, in the previous work for the BTZ black hole with two horizons, we have seen that the action variables for the two transition frequencies from the two families of the QNM lead to the quantization conditions of total horizon area and the difference between two horizon areas \cite{sk}. 

We can also consider the quantization of other black holes in this manner.  The three types of resonance modes can be obtained from the zeros and poles of transmission ($T$) and reflection ($R$) amplitudes for waves traveling from spatial infinity to the black hole horizon.  As an example, we consider  the Reissner-Nordstr{$\rm {\ddot {o}}$}m black hole. The  transmission and reflection probability in highly damped regime are given by  \cite{neitz} 
\begin{equation}
\fl \qquad T(\w) \tilde T(-\w)  \approx { {e^{{\w} \over T_H} -1} \over { {e^{{\w} \over T_H}} } +2+3e^{-{{\w} \over {T_{in}}}} } ~~ {\rm and}  ~~ R(\w) \tilde R(-\w)  \approx 3{ {1+e^{-{\w \over T_{in}}}} \over { {e^{{\w} \over T_H}} } +2+3e^{-{{\w} \over {T_{in}}}} }  ,
\end{equation}
where $T_{in} \equiv {\kappa_-} / (2 \pi)$ with the negative surface gravity $\kappa_-$ at inner horizon \cite{173}. 
These are for the scalar, electromagnetic and gravitational perturbations. 
From these, we find that the Reissner-Nordstr{$\rm {\ddot {o}}$}m black hole has three resonance modes  of QNM, TTM, and TRM in the highly damped regime. 
But, the highly damped QNM cannot be obtained algebraically from the poles of transmission $T$ and reflection $R$, while the highly damped  TTM and TRM are obtained as purely imaginary ones as follows:
\begin{eqnarray}
\w^{TTM} &=&  i 2 \pi  T_{in} \left( k+ {1/ 2} \right) , \\
 \w^{TRM} &=&  - i 2 \pi  T_{H}  k ,    \quad  ( k \in N ~{\rm and}~  k \gg 1 ) .
\end{eqnarray}
where $T_H$ is the Hawking temperature of the Reissner-Nordstr{$\rm {\ddot {o}}$}m black hole. 
Therefore  from the above TTM and TRM the transition frequencies  are obtained and  by applying the formula (\ref{myf}) the corresponding two action variables give the following quantization  conditions  via Bohr-Sommerfeld quantization:
\begin{eqnarray} 
 \I^{\pm} &=& M \sqrt{M^2-Q^2} \pm M^2 =  n_{\pm} \hbar , 
\end{eqnarray}
where the plus(minus) sign denotes TRM(TTM). 
It turns out that the TTM and TRM are associated with the quantizations of the inner horizon area and outer horizon area, respectively.  
It is easily found that the area and entropy spectra are given by $\triangle A_{out/in}=8 \pi  \hbar$ and $\triangle S=2 \pi$. 
Therefore we conclude that  the Schwarzschild, Kerr and Reissner-Nordstr{$\rm {\ddot {o}}$}m black holes have the universal behavior of  equally spaced entropy spectra as $\triangle S=2 \pi$. 
 The quantization of the inner horizon area can imply that there may be  physical dynamics inside the outer horizon.  For example, the Hawking radiation might happen at inner horizon \cite{make}. To clarify the physical meaning of the inner horizon spectrum, the further investigation on the physical dynamics inside outer horizon is needed.

\section{Conclusion}
We calculated the area and entropy spectra of the Kerr and Reissner-Nordstr{$\rm {\ddot {o}}$}m black holes. For this, we noticed that there are three types of highly damped resonance modes of black holes, which are  quasinormal modes (QNM),  total transmission modes (TTM) and total reflection modes (TRM). 
We proposed that all of these modes should be considered to carry information about quantum black hole since all they are  characteristic modes of  black hole, and therefore  they should be used in quantizing a black hole. 
Based on Bohr's correspondence principle, the quantum black hole with a transition frequency at large quantum number is considered as the classical periodic system with the oscillation frequency equal to the transition frequency in the semiclassical limit. The action variable $\I$  of the classical system of periodic motion is identified and quantized via the Bohr-Sommerfeld quantization in the semiclassical limit as the formula (\ref{myf}). 
We applied this method for the Kerr and Reissner-Nordstr{$\rm {\ddot {o}}$}m black holes. For the Kerr black hole, even though it was hard to obtain the quantization conditions from TTM and QNM, we could find  the  quantization condition from TRM.  
From this, we obtained that the outer horizon area spectrum of the Kerr black hole is equally spaced as $\triangle A_{out}=8 \pi  \hbar$,  consistent with Bekenstein's proposal \cite{be}. By Bekenstein-Hawking area law, the entropy spectrum also has equal spacing of $\triangle S=2 \pi$, which means that the Kerr black hole has the universal behavior of equally spaced entropy spectrum like other black holes in Refs.\cite{sk, sk2}. 
For the Reissner-Nordstr{$\rm {\ddot {o}}$}m black hole, we found that the quantization conditions from TRM and TTM lead to the quantization of the outer and inner horizon area, respectively. Since the area spectra are equally spaced as $\triangle A_{out/in}=8 \pi  \hbar$, we also found that the  Reissner-Nordstr{$\rm {\ddot {o}}$}m black hole has the universal behavior of  equally spaced entropy spectrum as $\triangle S=2 \pi$. 
These results agree with the quantization of the entropy spectrum obtained in different methods of Refs.\cite{pad, ropo, med2}. 
Our results also give good examples for the claim in  \cite{sk2} that there is the universality that the entropy spectrum of a black hole is equally spaced. Therefore  we found that the universality holds regardless of the dimension of spacetime, the presence of the angular momentum or charge, and the gravity theory. 
It is expected that the universal behavior of entropy spectrum would be useful for understanding and investigating a quantum nature of black holes as the first step toward quantum gravity. 
%
%
%
%

%
%
\ack{ YK thanks Dr.Jong-Dae~Park for useful discussions. This research  was supported by the National Research Foundation of Korea(NRF) grant funded by the Korea government(MEST)(No.2009-0063068) and  also supported by Basic Science Research Program through the National Research Foundation of Korea(NRF) funded by the Ministry of Education, Science and Technology(No.2010-0008109).}
%
%
%


\section*{References}
\begin{thebibliography}{99}

\bibitem{be} 
Bekenstein J D 1974,  (1974) 
{\it Lett. Nuovo Cim.} {\bf 11}  467
; Bekenstein J D 1997 ({\it \pre} arXiv:gr-qc/9710076)    

\bibitem{1302}
 Corichi A 2009,  ({\it \pre} arXiv:0901.1302)  and references therein.     


\bibitem{122dryer}
Corichi A, Diaz-Polo J, and Fernandez-Borja E 2007, {\it Phys. Rev. Lett.} {\bf 98} 181301  
  ({\it \pre} gr-qc/0609122) 
; Dreyer O 2003, {\it Phys. Rev. Lett.} {\bf 90} 081301   
   ({\it \pre} gr-qc/0211076) 

\bibitem{hod} 
Hod S 1998, {\it Phys. Rev. Lett.} {\bf 81} 4293  
({\it \pre} gr-qc/9812002)

\bibitem{kun} 
Kunstatter G 2003, {\it Phys. Rev. Lett.} {\bf 90} 161301  
 ({\it \pre} gr-qc/0212014)

\bibitem{setvag}  
Setare M R and  Vagenas E C 2005, {\it Mod. Phys. Lett.} A {\bf 20} 1923  
  ({\it \pre} hep-th/0401187)

\bibitem{magi} 
Maggiore M 2008, {\it Phys. Rev. Lett.} {\bf 100} 141301 
 ({\it \pre} arXiv:0711.3145)


\bibitem{vage} 
Vagenas E C 2008,  {\it  JHEP} {\bf 11} 073  
({\it \pre} arXiv:0804.3264) 

\bibitem{med} 
Medved A J M 2008, {\it Class. Quant. Grav.} {\bf 25} 205014  
({\it \pre} arXiv:0804.4346)

\bibitem{wei}
Wei S-W, Li R, Liu Y, and Ren J 2009, {\it JHEP} {\bf 03} 076 
 ({\it \pre} arXiv:0901.0587)


\bibitem{wei2}
Wei S and Liu Y, 
 ({\it \pre} arXiv:0906.0908) 
 

\bibitem{020}
Andersson N and Howls C J 2004, {\it  Class. Quant. Grav.} {\bf 21} 1623    
 ({\it \pre} gr-qc/0307020) 

\bibitem{173}
Motl L and Neitzke A 2003, {\it  Adv. Theor. Math. Phys.} {\bf 7} 307   
 ({\it \pre}   hep-th/0301173) 

\bibitem{029}
Berti E and Kokkotas K D 2003, {\it Phys. Rev. D} {\bf 68} 044027   
 ({\it \pre} hep-th/0303029) 

\bibitem{052}
Berti E, Cardoso V, and Yoshida S 2004, {\it Phys. Rev. D} {\bf 69} 124018   
 ({\it \pre} gr-qc/0401052) 

\bibitem{keshk}
Keshet U and  Hod S 2007, {\it Phys. Rev. D} {\bf 76} 061501  
({\it \pre} arXiv:0705.1179) 

\bibitem{keshk2}
Keshet U and Neitzke A 2008, {\it Phys. Rev. D} {\bf 78} 044006 
 ({\it \pre} arXiv:0709.1532)

\bibitem{4195}
Kao H-C and Tomino D 2008, {\it Phys. Rev. D} {\bf 77} 127503   
 ({\it \pre} arXiv:0801.4195) 

\bibitem{myung} 
Myung Y S 2010, {\it Phys. Lett. B} {\bf 689} 42 
 ({\it \pre} arXiv:1003.3519) 

\bibitem{sk}
Kwon Y and Nam S 2010, {\it Class. Quant. Grav.} {\bf 27} 125007  
({\it \pre} arXiv:1001.5106)

\bibitem{sk2}
Kwon Y and Nam S 2010, {\it Class. Quant. Grav.} {\bf 27} 165011   
({\it \pre} arXiv:1002.0911)

\bibitem{chan}
Chandrasekhar S 1984, {\it Proc. Roy. Soc. Lond.} A {\bf 392} 1 

\bibitem{ono}
Onozawa H 1997, {\it  Phys. Rev. D} {\bf 55} 3593  
({\it \pre} gr-qc/9610048) 

\bibitem{van}
Maassen van den Brink A 2000, {\it Phys. Rev. D} {\bf 62} 064009   
 ({\it \pre}  gr-qc/0001032) 

\bibitem{4306}
Dotti G,  Gleiser R J, Ranea-Sandoval I F, and Vucetich H 2008,  {\it Class. Quant. Grav.} {\bf 25} 245012   
 ({\it \pre}  arXiv:0805.4306)  

\bibitem{29751025}
Berti E, Cardoso V, and  Starinets A O 2009, {\it Class. Quant. Grav.} {\bf 26} 163001 
({\it \pre}  arXiv:0905.2975)  
; Berti E 2004 ({\it \pre}  gr-qc/0411025 )  

\bibitem{nolkz}
Nollert H-P 1999, {\it Class. Quant. Grav.} {\bf 16}  R 159  ; 
Kokkotas K D and Schmidt B G 1999, {\it Living Rev. Rel.} {\bf 2} 2  

\bibitem{area0}
 Hawking S W 1975, {\it Commun. Math. Phys.} {\bf 43} 199   

\bibitem{neitz}
Neitzke A 2003, ({\it \pre} hep-th/0304080 )      

\bibitem{teu}
Teukolsky S A 1972, {\it Phys. Rev. Lett.} {\bf 29} 1114  
;  Teukolsky S A 1973, {\it Astrophys. J.} {\bf 185} 635  

\bibitem{area}
Bekenstein J D 1973, {\it Phys. Rev. D} {\bf 7} 2333  
 ; Bekenstein J D 1972, {\it Lett. Nuovo Cim.} {\bf 4} 737  
; Hawking S W 1974, {\it Nature} {\bf 248} 30  


\bibitem{muin}
M-I. Park, 
 Phys. Lett. B {\bf 647} 472 (2007)  ({\it \pre}  hep-th/0602114 )  


\bibitem{make}
Makela J, Repo P, Luomajoki M, and Piilonen J 2001, 
{\it  Phys. Rev. D} {\bf 64}  024018   ({\it \pre} gr-qc/0012055)

\bibitem{pad}
Kothawala D, Padmanabhan T, and Sarkar S 2008, {\it Phys. Rev. D} {\bf 78} 104018  
({\it \pre} arXiv:0807.1481) 

\bibitem{ropo}
Ropotenko K 2009, {\it Phys. Rev. D} {\bf 80}  044022  
({\it \pre} arXiv:0906.1949) 

\bibitem{med2}
Medved A J M 2009, {\it Mod. Phys. Lett.} A {\bf 24} 2601  
 ({\it \pre} arXiv:0906.2641) 



 




\end{thebibliography}
\end{document}